\documentclass[12pt]{iopart}

\begin{document}

\title[Example of pressure excitation]{A simple example of pressure excitation:the 3D square well potential}

\author{V.Celebonovic and M.G.Nikolic}

\address{Institute of Physics,Univ.of Belgrade,Pregrevica 118,11080 Belgrade,Serbia}
\ead{vladan@ipb.ac.rs}
\vspace{10pt}
\begin{indented}
\item[]May 2016
\end{indented}

\begin{abstract}


High pressure experiments in diamond anvil cells as well as in geophysics show that at certain discrete values of pressure abrupt changes of the mass density and changes of the material structure occur. The aim of this paper is to discuss  the influence of high external pressure on atoms and molecules using the $3D$ square well potential as a simple solvable quantum mechanical example. Values of external pressure needed for a transition between energy levels with different values of the main quantum number $n$, as well as the pressure needed to expel the particle from this system are calculated. In both cases discrete values of pressure are obtained.
\end{abstract}

%
\vspace{2pc}
\vspace{2pc}

\maketitle
\normalsize
%
%
%
%
%
%
%
%

\section{Introduction}

The only correct theoretical approach to studies of atoms and  molecules is the use of quantum mechanics. The essence of such studies is finding solutions of the Schr\"{o}dinger equation, and thus determining the energies and wave functions of particles in the problem. Such studies are rendered more complex by the introduction of the influence of external fields in the calculation. The external fields encountered most frequently in this kind of research are the electrical and magnetic fields.

An equally interesting but perhaps more exotic case of an external field influencing the energy levels of a bound particle is the situation in which an atom and/or molecule is subdued to high external pressure. Studies of the behavior of materials under high external pressure are considered as "exotic" by terrestrial standards, but such conditions are frequent in geophysics and astronomy, for example in planetary interiors. 

Various studies (recent examples are \cite{da81},\cite{du05}) have led to the conclusion that discontinous changes of density occur at the mutual boundaries of layers in the interior of the Earth.Similar changes have been detected in diamond anvil cell (DAC) experiments such as \cite{grx13}, so the obvious question is the possibility of theoretically predicting the value of external pressure needed for the occurrence of such discontinuous changes. This value of external pressure will correspond to the difference of values of the internal pressure corresponding to two energy levels of the atom or molecule. First attempts of determining the internal pressure in atoms were made early in the last century (for example, \cite{r23},\cite{br35}).

According to \cite{as04}  Bridgman was the first to propose that "the stable structure of atoms could be broken down in the condensed state of matter by mutual encroachment of atoms, this being brought about by pressure". Older literature (such as \cite{sch58}) mentions that Fermi was the first to attempt solving the Schr\"{o}dinger equation for atoms under high pressure. It seems that he just noted that external pressure induces changes in atomic structure, but that he had no interest in elaborating any further. 
 
The first suggestion of a possibility of theoretically studying the effects of high external pressure on a hydrogen atom was given by \cite{mi37},followed by \cite{sw38}. Paper \cite{sw38} was neglected for a long time, due to lack of clear experimental motivations for such calculations.  More recently,a quantum mechanical theory of the influence of external pressure on atoms and molecules in a real solid was developed for the explanation of the pressure shift of the spectral lines of ruby \cite{ma88}. 

The aim of this paper is to discuss the behavior under high external pressure of a well known quantum mechanical system - the $3D$ square well potential . This system is analytically solvable, and it can be considered as a simplified model of a particle bounded by a central potential. An expression for the value of external pressure needed for excitation, and ultimately ionisation of such a system will be derived. This result will represent an approximate estimate of the pressure needed to excite a bound particle. The simplest real physical system which could be described by this model is the hydrogen atom, whose behavior under high pressure is a subject of intensive studies for the last 80 years.

It could be objected at this point that pressure is a macroscopic quantity, applicable to many body systems, and that it would be contrary to the principles of statistical mechanics to apply it to an atom. According to \cite{ba94},\cite{ba97} and references given there, the pressure can be introduced as a mechanical property of an atom by identifying the atom with a proper open system - $POS$. A $POS$ is a region of real space whose properties are defined by the principle of stationary action. It provides a variational statement of the Heisenberg equation of motion for the product of the position and momentum operators, and thus the virial theorem of a $POS$. A $POS$ is bounded by a surface of zero flux of the vector field $\nabla\rho(r)$,where $\rho(r)$ is the electron density. The virial of such a system consists of two parts: the virial of the forces acting within the system and of the forces acting on its surface. The second term is proportional to the pressure - volume product, and it is used to define the pressure.

Apart the introduction, this paper has two more sections. The next one is devoted to a brief reminder of standard quantum mechanical results on this system, while the third part discusses the calculation of the pressure needed for a transition of a particle  between two arbitrary energy levels of a $3D$ square potential well.

\section{Some quantum mechanics}

The three dimensional square well potential is defined by the relation \cite{sch68}

\begin{eqnarray}
\ V(r) = \left\{ 
\begin{array} {c l}
- V_{0} & r <a\\
0 &  r >a
\end{array}\right.  
\end{eqnarray}
where the symbol $a$ denotes the radius of the well. The volume of this system is $V=\pi a^{2} V_{0}$, which has the dimensions of surface multiplied by the energy. Introduce the following  definitions
\begin{equation}
\alpha^{2} = \frac{2 m}{\hbar^{2}} (V_{0}-|E|) ; \beta^{2} = \frac{2 m}{\hbar^{2}} |E|
\end{equation} 

and the change of variables

\begin{equation}
\xi = \alpha a ; \eta = \beta a
\end{equation}

which leads to
\begin{equation}
\xi^{2}+\eta^{2} = \frac{2m}{\hbar^{2}} V_{0} a^{2}
\end{equation} 
This equation represents a circle in the $(\xi,\eta)$ plane with a radius 
equal to the value
$(2 m V_{0} a^{2}/\hbar^{2})^{1/2}$. The left hand side of eq. (4) is dimensionless, which 
\begin{equation}
 \gamma = \frac{2mV_{0}}{\hbar^{2}} = \frac{n}{a^2}
\end{equation}
where $n$ is an integer.
It can be shown (for example \cite{sch68}) 
that for energy levels with $l= 0$,where $l$ denotes the angular momentum quantum number, 
\begin{equation}
\eta = - \xi\cot[\xi]
\end{equation}
The functions $\xi$ and $\eta$ are positive by definition. This means that the energy levels are given as intersection points of curves represented by equations (4) and (6) in the region $\xi\geq0$, $\eta\geq0$. Using these expressions , it follows that the intersection points are given by:

\begin{equation}
\sqrt{n-\xi^{2}} + \xi\cot(\xi) = 0
\end{equation}
 
In order for $\xi$ to be real and positive, it follows that eq.(7) is fulfilled ( that is an energy level exists) if $\xi^{2} < n$ and
$\pi >\xi > \pi/2$ .


In the case of $l=1$, the position of the energy levels is given by the following equations \cite{sch68}: 
\begin{eqnarray}
\frac{\cot[\xi]}{\xi} - \frac{1}{\xi^{2}} = \frac{1}{\eta}+\frac{1}{\eta^{2}} \nonumber\\ 
\xi^{2} + \eta^{2} = \frac{2m}{\hbar^{2}} V_{0} a^{2}
\end{eqnarray}
The method of solving this system of equations is discussed in \cite{sch68}.  

\section{The influence of external pressure} 
Pressure is defined as the negative volume derivative of the energy of the system, that is
\begin{equation}
P = - \frac{\partial E}{\partial V}
\end{equation}
or 
\begin{equation}
P = - \frac{\partial E}{\partial V}= -\frac{\partial E}{\partial \xi}\frac{\partial \xi}{\partial n} \frac{\partial n}{\partial V} 
\end{equation}


Definitions of $\alpha$ and $\xi$ and eq.(5), applied to a bound state  arbitrarily denoted by $1$ lead to
\begin{equation}
|E_{1}| = V_{0} (1-\frac{\xi_{1}^{2}}{n_{1}})
\end{equation}
which implies that $\xi^{2} < n$ and means that
\begin{equation}
\frac{\partial |E_{1}|}{\partial \xi_{1}} = - 2V_{0} \frac{\xi_{1}}{n_{1}} \theta(n-n_{1}) 
\end{equation}

where $\theta$ denotes the Heaviside step function. As 

\begin{equation}
\xi_{1} = \alpha_{1} a_{1} = \alpha_{1} \sqrt{\frac{n_{1}}{\gamma_{1}}}
\end{equation}
it follows that
\begin{equation}
\frac{\partial \xi_{1}}{\partial n_{1}} = \frac{\alpha_{1}}{2 \sqrt{\gamma_{1} n_{1}}} \theta(n-n_{1}) 
\end{equation}
Finally,
\begin{equation}
n_{1} = \gamma_{1} a_{1}^{2} = \frac{2 m}{\pi \hbar^{2}} V_{1} 
\end{equation} 
Therefore 
\begin{equation}
\frac{\partial n_{1}}{\partial V_{1}} = \frac{2 m}{\pi \hbar^{2} } \theta(V-V_{1}) 
\end{equation}
Combining eqs. (10), (12),(14) and (16) gives the following expression for the internal pressure corresponding to the energy level $n_{1}$ of a $3D$ finite square well potential 
\begin{eqnarray}
P[n_{1}] = \frac{2 m V_{0}}{\pi \hbar^{2}} \frac{\alpha_{1}^{2} a}{\gamma_{1}^{1/2} n_{1}^{3/2}}\times\nonumber\\
\theta(n-n_{1})\theta(n-n_{1}) \theta(V-V_{1})    
\end{eqnarray}
which after some algebra can be transformed into the following final form
\begin{equation}
P[n_{1}] = \frac{2 m V_{0}}{\pi \hbar^{2}} \frac{\xi_{1}^{2}}{n_{1}^{2}}
\end{equation}
This is the final expression for the internal pressure at the energy level having the "main quantum number" $n_{1}$ in a $3D$ square well potential . The quantity which is of experimental relevance is the external pressure needed for a transition between levels with quantum numbers $n_{1}$ and $n_{2}$, and which is equal to the difference $P[n_{1}]$-$P[n_{2}]$. It can be shown that 
\begin{equation}
P_{tr} = \frac{2 m V_{0}}{\pi \hbar^{2}}\times\frac{\xi_{1}^{2}}{n_{1}^{2}} \times[1-(\frac{n_{1}}{n_{2}})^{2} (\frac{\xi_{2}}{\xi_{1}})^{2}]
\end{equation}
 As $P_{tr}$ is a measurable quantity, it must be positive, which implies that the following condition must be fulfilled for a transition to be physically possible: 
\begin{equation}
[\frac{\xi_{2}}{\xi_{1}} \frac{n_{1}}{n_{2}}]^{2}\leq1
\end{equation}
Reverting explicitly to the energies of the levels between which a transition occurs, the following final expression is obtained for the transition pressure
\begin{equation}
P_{tr} = \frac{2 m V_{0} }{\pi \hbar^{2}}\frac{1}{n_{1}} (1-(\frac{|E_{1}|}{V_{0}}))\times[1-(\frac{n_{1}}{n_{2}})(\frac{(V_{0}-|E_{2}|)}{V_{0}-|E_{1}|})]
\end{equation}
The pressure needed to ionize a $3D$ square well potential in which a particle is in a state with quantum number $n$ can be estimated from eq.(21). This system will become ionized if $|E_{2}|\rightarrow V_{0}$ and $n_{2}\rightarrow\infty$. 

Inserting these two conditions into eq.(21) it follows that the external pressure needed to ionize a $3D$ square well potential containing a particle with the main quantum number $n_{1}$  is given by

\begin{equation}
P_{ion} = \frac{2 m V_{0}}{\pi \hbar^{2}} \frac{1}{n_{1}} \frac{V_{0}-|E_{1}|}{V_{0}}
\end{equation}

\subsection{Some numerical estimates}
Solving eq.(7)  one can obtain values of $\xi$ corresponding to various $n$. The value of external pressure needed for a transition between energy levels with $n_{1}$ and $n_{2}$ is given by eq.(21). As two arbitrarily chosen examples, take the transitions  $3\rightarrow4$ and $4\rightarrow5$. It can be shown that $P_{3\rightarrow4}=0.204$ $m V_{0}/\hbar^{2}$. Proceeding in the same way, one gets that $P_{4\rightarrow5}=0.124$ $m V_{0}/\hbar^{2}$.

The value of external pressure needed for ionizing a $3D$ square well potential was calculated for two arbitrary values of $n$. It turned out that $P_{ion}=0.653$ $mV_{0}/\pi\hbar^{2}$ in the case $n=3 $ and $P_{ion}= 0.4931$ $m V_{0}/\pi\hbar^{2}$ in the case  $n= 6$.
\section{Discussion and conclusions}
The aim of this paper was to calculate the value of external pressure needed for excitation and ionisation of a $3D$ square well potential. The results are given by eqs.(21) and (22). The calculations leading to these results are mathematically simple, but they enable some interesting conclusions to be drawn. 
The value of pressure obtained in eq.(21) depends on the energies and therefore on the main quantum numbers of the two energy levels between which the transition occurs. Both the energies and the quantum numbers are discrete, which implies that the transition between any two energy levels in a $3D$ square well potential is possible only for certain discrete values of pressure. Discrete values of the pressure imply discontinous changes of the density, which is in agreement with experimental results.  Various physically interesting limiting cases are easily visible from eq.(21). 

If $|E_{2}|\rightarrow V_{0}$ , which means that the particle under consideration is near the top of the potential well, one gets eq.(22), that is the pressure needed for ionization. Another interesting limiting case is $n_{2}>>n_{1}$. Physically, this corresponds to a situation in which the energy level to which a transition occurs is much higher in the energy scale than the energy level from which the transition starts. Such a situation is approximately equal to ionisation, so the required external pressure is approximately given by eq.(22). 

Discontinuous changes of material parameters are often expressed in terms of mass densities at which they occur,while in this paper we have obtained values of external pressure needed for such changes, within a particular simple model. The conversion could be performed by using an appropriate form of the equation of state for the material under study. For a detailed review of the problems related to equation of state of solids see for example \cite{st05} and later work. 

Take for example the so called Birch-Murnaghan equation of state. Its standard form is
\begin{equation}
P(\rho)=\frac{3 B_{0}}{2} [1+\frac{3}{4}(B_{0}'-4)(x^{2/3}-1)][x^{7/3}-x^{5/3}]
\end{equation}
where $x=\rho/\rho_{0}$, $B_{0}$ is the bulk modulus of the material and  $B_{0}'$ is the pressure derivative of $B_{0}$. Let $P_{1},\rho_{1}$, $P_{2},\rho_{2}$ denote values of the pressure and density at two points of the $P-\rho$ phase diagram. Introduce the difference of the two values of the density by $\Delta = \rho_{2}-\rho_{1}$.
Using eq.(23) to form an expression for $P_{2}-P_{1}$ and expanding, one gets the following approximate result
\begin{equation}
P_{2}-P_{1} \cong (\frac{B_{0}'}{4}-1)\times[\frac{9B_{0}}{2} (\Delta/\rho_{0})^{3}+\frac{27 B_{0}}{2} (\rho_{1}/\rho_{0})(\Delta/\rho_{0})^{2}]+...
\end{equation}
If $P_{2}-P_{1}=P_{tr}$ it is clear hat the last expression gives a link between measurable quantities and the main quantum numbers of the energy levels between which a transition occurs.

The calculation discussed in this letter was performed for two values of the main quantum number $n$. 
As the energy level with a higher value of $n$ has a smaller binding energy, it is physically plausible that the external pressure needed to ionise a level with a higher value of $n$ is smaller than the corresponding value for the level with a lower value of $n$. 

The results of this paper illustrate the idea that only certain discrete values of external pressure
can provoke excitation and ionisation at the atomic/molecular level thus leading to phase transitions at the 
"macroscopic"level.  It is planned to perform a similar calculation for the Lennard - Jones potential, which is 
more realistic and for which the Schr\"{o}dinger equation has been solved only recently \cite{ses13}.

\ack
This paper was prepared within the research projects $OI171017$ and $171038$ financed by the Ministry of Education,Science and Technological Development of Serbia. V.C. is grateful to  Prof.S.Fassari for useful correspondence and for pointing out ref \cite{ses13}.


\end{document}